# Applied Surface Science
## Multi-functional oxidase-like activity of praseodymia nanorods and nanoparticles
--Manuscript Draft--

| | |
|---|---|
| Manuscript Number: | APSUSC-D-22-10755R2 |
| Article Type: | Full Length Article |
| Keywords: | praseodymia; Nanorods; nanoparticles; artificial enzyme; oxidase |
| Corresponding Author: | Xiaowei Chen, Ph.D.<br>University of Cadiz<br>Puerto Real, Cadiz SPAIN |
| First Author: | Lei Jiang, Ph.D. |
| Order of Authors: | Lei Jiang, Ph.D. |
| | Yaning Han |
| | Susana Fernández-García, Ph.D. |
| | Miguel Tinoco, Ph.D. |
| | Zhuang Li |
| | Pengli Nan |
| | Jingtao Sun |
| | Juan Delgado, Ph.D. |
| | Huiyan Pan, Ph.D. |
| | Ginesa Blanco, Ph.D. |
| | Javier Martínez-López, Ph.D. |
| | Ana Hungria, Ph.D. |
| | Jose Calvino, Ph.D. |
| | Xiaowei Chen, Ph.D. |
| Abstract: | The ability to mimic protein-based oxidase with multi-functional inorganic nanozymes would greatly advance biomedical and clinical practices. Praseodymia (PrO $_x$ ) nanorods (NRs) and nanoparticles (NPs) have been synthesized using hydrothermal and precipitation methods. Both PrO $_x$ catalysts with different morphologies exhibit significantly higher oxidase-like activities (Michaelis-Menten constant $K_m$ £ 0.026 mM) than commercial PrO $_x$ and most so-far-reported artificial enzymes. One of the substrates, dopamine, can be oxidized and further polymerized to generate polydopamine in acidic conditions. Akin to CeO $_2$ , which is a well-studied nanozyme, a different mechanism involving holes $^+$ , oxygen vacancies and oxygen mobility over PrO $_x$ catalysts has been proposed in this work. However, fluoride ions were found to impose opposite effects on the oxidase-mimicking activity of PrO $_x$ and CeO $_2$ , implying a promising path for the exploration of new nanozymes. In support of this, PrO $_x$ was further applied in colorimetric sensing of L-cysteine and fluoride with high sensitivity. |
| Suggested Reviewers: | Mengfei Luo, PhD<br>Zhejiang Normal University<br>mengfeiluo@zjnu.cn<br>Professor Luo has been working on lanthanide-based catalysts for a long time. |
| | Mingyuan Zheng, PhD<br>Dalian Institute of Chemical Physics Chinese Academy of Sciences<br>myzheng@dicp.ac.cn<br>Professor Zheng has been working on heterogeneous catalysis. |
| | Juewen Liu, PhD<br>University of Waterloo |

| | liujw@uwaterloo.ca |
|---|---|

# Highlights

- Synthesis of praseodymia nanorods and nanoparticles
- Excellent multi-funtional oxidase-like activities of praseodymia nanomaterials
- Colorimetric sensing of L-cysteine and fluoride using praseodymia nanorods

## Graphical abstract

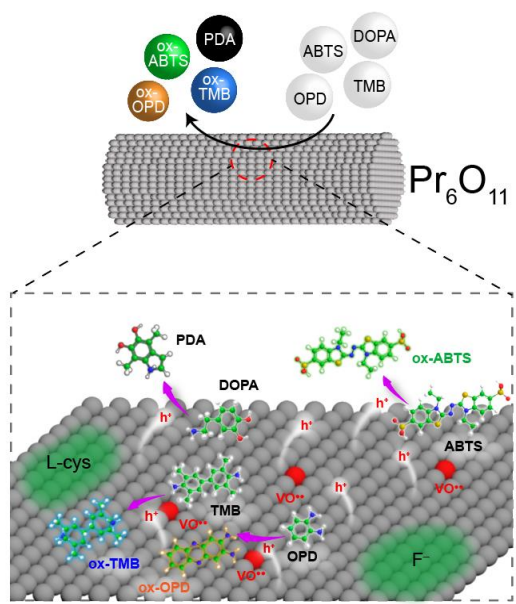

# Multi-functional oxidase-like activity of praseodymia nanorods and nanoparticles

Lei Jiang [a, *], Yaning Han [a], Susana Fernández-García [b], Miguel Tinoco [b, c], Zhuang Li [a], Pengli Nan [a], Jingtao Sun [a], Juan J. Delgado [b, e], Huiyan Pan [b, d], Ginesa Blanco [b, e], Javier Martínez-López [f], Ana B. Hungría [b, e], Jose J. Calvino [b, e], Xiaowei Chen [b, e, *]

[a] Heavy Oil State Laboratory and Center for Bioengineering and Biotechnology, China University of Petroleum (East China), Qingdao, 266580, China

[b] Departamento de Ciencia de los Materiales, Ingeniería Metalúrgica y Química Inorgánica, Facultad de Ciencias, Universidad de Cádiz, Campus Río San Pedro, Puerto Real (Cádiz), E-11510, Spain

[c] Departamento de Química Inorgánica, Facultad de Ciencias Químicas, Universidad Complutense de Madrid, Av. Complutense S/N, Madrid (Madrid) E-28040, Spain

[d] Henan Key Laboratory of Industrial Microbial Resources and Fermentation Technology, College of Biological and Chemical Engineering, Nanyang Institute of Technology, Nanyang, 473004, China

[e] Instituto Universitario de Investigación en Microscopía Electrónica y Materiales (IMEYMAT), Universidad de Cádiz, Campus Río San Pedro, Puerto Real (Cádiz), E-11510, Spain

[f] Departamento de Ciencias de la Tierra, Universidad de Cádiz, Campus Río San Pedro, Puerto Real (Cádiz), E-11510, Spain

* Corresponding authors: L. Jiang, tel: 0086-532-86981568, fax: 0086-532-86981569, Email address: leijiang@upc.edu.cn; X. Chen, tel: 0034-956-012741, Email address: xiaowei.chen@uca.es;

**Abstract**

The ability to mimic protein-based oxidase with multi-functional inorganic nanozymes would greatly advance biomedical and clinical practices. Praseodymia ($PrO_x$) nanorods (NRs) and nanoparticles (NPs) have been synthesized using hydrothermal and precipitation methods. Both $PrO_x$ catalysts with different morphologies exhibit significantly higher oxidase-like activities (Michaelis-Menten constant $K_m \leq 0.026$ mM) than commercial $PrO_x$ and most so-far-reported artificial enzymes. One of the substrates, dopamine, can be oxidized and further polymerized to generate polydopamine in acidic conditions. Akin to $CeO_2$, which is a well-studied nanozyme, a different mechanism involving holes$^+$, oxygen vacancies and oxygen mobility over $PrO_x$ catalysts has been proposed in this work. However, fluoride ions were found to impose opposite effects on the oxidase-mimicking activity of $PrO_x$ and $CeO_2$, implying a promising path for the exploration of new nanozymes. In support of this, $PrO_x$ was further applied in colorimetric sensing of L-cysteine and fluoride with high sensitivity.

**Keywords:** praseodymia, nanorods, nanoparticles, artificial enzyme, oxidase

**1. Introduction**

Oxidase is one of the most important multi-functional enzyme family in nature. This kind of enzymes has ability to oxidize chemicals containing groups such as amino, polyphenol or sulfur [1]. Mimicking or replacing protein-based oxidase would significantly advance both biomedical research and biochemical applications. As a means to an end, recent studies have shown the development of many attractive oxidase-mimicking nanozymes involving noble metals (Pt, Au and Ag) [2] or metal oxides ($Fe_3O_4$, $CeO_2$, CuO, $MnO_2$ and $Mn_3O_4$) [2]. A representative nanozyme is $CeO_2$, a rare earth oxide (REO) with excellent oxygen storage capability, rich surface charges, and thus, multiple oxidase-like properties [3]. Lately, Chen et al. reported the so-far most active oxide-mimicking $Mn_2O_3$ (surface area 8.5 $m^2/g$) with kinetic parameter $K_m$ of 0.13 mM [4]. Despite the tremendous efforts, the exploration of nanozymes with high catalytic activity and selectivity remains a great challenge.

Praseodymium is the neighbor of cerium in the periodic table of elements. $CeO_2$ is the predominant stoichiometry of cerium oxide, while the common phase of praseodymium oxide under ambient condition is roughly $Pr_6O_{11}$ [5, 6]. $Pr_6O_{11}$ can be considered as an oxygen-deficient modification of cubic fluorite-like $PrO_2$ and contains already one third of Pr in its trivalent state and two thirds of $Pr^{4+}$. $Pr_6O_{11}$ has been reported to have the highest oxygen mobility of all undoped REOs [7]. From a doping point of view, it can be attributed to the fact that $Pr^{4+}$ is naturally "doped" with $Pr^{3+}$ in $Pr_6O_{11}$. In this sense, praseodymia has a promising potential to be a catalytic artificial enzyme because of its $Pr^{3+}/Pr^{4+}$ pair. It has been used as catalyst additive or support in catalytic oxidation of CO, methanol, methane [8] and soot [9], decomposition reactions of organic molecules such as 2-propanol, oxidative coupling of methane and liquid-phase benzylation. For soot oxidation, three-dimensionally ordered macro-porous praseodymia was significantly more active than ceria [9]. Our previous studies have shown that Pr-doped ceria nanocubes are very active as artificial enzymes, from oxidant (oxidase-like) to antioxidant (hydroxyl radical scavenging), as well as paraoxon degradation (phosphatase-like) activities [3, 10]. However, application of praseodymia as an oxidase has not been ever studied yet.

The preparation method can strongly affect textural, structural and redox features of praseodymia, such as phase composition, surface area, particle morphology or size and $Pr^{3+}/Pr^{4+}$ ratio on surface. These properties subsequently tune their catalytic activities as nanozymes. Both preseodymia NRs and NPs can be synthesized via precipitation, hydrothermal, solvothermal and microwave-assisted methods [11], after thermal annealing of $Pr(OH)_3$ or thermal decomposition of praseomymium compounds [12]. Despite $Pr_6O_{11}$ is one of well-known praseodymia phases, it is true that $Pr^{3+}/Pr^{4+}$ ratio can change easily, depending on the preparation or storage conditions of the oxide [13]. For this reason, we will henceforth use the $PrO_x$ formula to refer praseodymia.

In this work, $PrO_x$ NRs and NPs, with relatively high surface areas, have been synthesized using hydrothermal and precipitation methods. Their oxidase-like activities for oxidation of 3,3',5,5'-tetramethylbenzidine (TMB), 2,2-azinobis-(3-ethylbenzothizoline-6-sulfonic acid) (ABTS), o-phenylenediamine (OPD), and dopamine (DOPA) have been studied using commercial praseodymia as a reference sample. The mechanism and key intermediates, as well as the effects of anions and amino acids have been investigated to use $PrO_x$ as a new nanozyme and promote the development of biomarker sensors for L-cysteine and $F^-$ anions.

**2. Materials and Methods**

*2.1. Synthesis of $PrO_x$ NRs and NPs*

A hydrothermal method was used to synthesize $PrO_x$ NRs ($PrO_x$-NR) [11, 14]. 240 mL mixture of 6 M NaOH (Alfa Aesar, 98%) and 0.05 M $Pr(NO_3)_3 \cdot 6H_2O$ (Aldrich, 99.9%) aqueous solution was stirred in a 300 mL Teflon container for 30 min. Then, this container was sealed and placed in a stainless steel autoclave. The autoclave was heated to 180 ºC and maintained at this temperature for 24 h. Afterwards, the autoclave was cooled down to room temperature and the resulting suspension was centrifuged and washed with deionized water several times and then with ethanol once (Panreac, Absolute Ethanol). Finally, sample was dried at 80 ºC for 24 h in an oven and calcined at 500 ºC for 4 h in a muffle furnace.

For the synthesis of $PrO_x$ NPs ($PrO_x$-NP), a precipitation method was used [15]. An aqueous solution of 0.067 M $Pr(NO_3)_3 \cdot 6H_2O$ (Alfa Aesar, 99.9%) was mixed with 4.9 M $NH_3 \cdot H_2O$ in 75 mL solution and stirred at 80 ºC for 3 h. The mixture was kept at room temperature for 48 h followed by centrifugation for 30 min. The obtained precipitate was rinsed with ultrapure water and ethanol over three times, dried at 100 ºC in an oven for 10 h, and finally calcined at 500 ºC for 3 h in a muffle furnace.

A commercial $Pr_6O_{11}$ (99.99%) sample ($PrO_x$-C) was purchased from Alfa Aesar.

*2.2. Physical and compositional characterization*

X-Ray diffraction (XRD) patterns were recorded on a D8 ADVANCE diffractometer of Bruker using a Cu Kα radiation, with a range of 25-75º, a step of 0.02º and a step time of 1 s. Phase identification and composition calculation were performed using TOPAS v5 software from Bruker. Brunauer-Emmett-Teller (BET) surface areas of samples were determined in a Micromeritics TriStar - 3020 via $N_2$ adsorption at -196.15 ºC. Prior to analysis, samples were degasified at 150 ºC for 8 h under vacuum.

The surface chemical composition and oxidation states of samples were characterized by X-ray photoelectron spectroscopy (XPS). Analyses were performed on an ESCALAB 250Xi instrument. Spectra were recorded using monochromatized Al Kα X-Ray (1486.6 eV), with an X-ray power of 150 W. The spectrometer was operated in a constant analyzer energy mode, with pass energy of 30 eV. Powder samples were pressed into a disk, which was stuck on a double-sided adhesive conducting polymer tape and analyzed without further treatment. The binding energy scale was calibrated with respect to the C *1s* signal at 284.8 eV. CasaXPS software, version 2.3.23.PR1.0 (Casa Software Ltd., Devon, UK), was used for spectra processing.

The morphology and size of $PrO_x$ samples were studied by Scanning Electron Microscopy (SEM). A FEI Nova Nano SEM450 microscope was used at 3 to 10 kV accelerating voltages with a working distance of 5 mm. $PrO_x$ samples were also characterized by High Resolution Transmission Electron Microscopy (HRTEM) and Scanning Transmission Electron Microscopy-High Angle Annular Dark Field imaging (STEM-HAADF) using JEOL 2010-F and FEI Titan[3] Themis 60–300 aberration-corrected microscopes. HRTEM images were obtained with 0.19 nm spatial resolution at Scherzer defocus and STEM-HAADF images were acquired by using an electron probe of 0.5 nm of diameter at a camera length of 8 cm.

Temperature Programmed Reduction with $H_2$ ($H_2$-TPR) started with a pretreatment consisting in an oxidation under a 5% $O_2$/He flow (60 mL/min) at 500 ºC for 1 h. After the oxidation pretreatment, samples were cooled in the same 5% $O_2$/He flow down to 150 ºC, and then, the flow was switched to pure He cooling down to room temperature. The samples were reduced in a 60 mL/min flow of 5% $H_2$/Ar with a heating rate of 10 ºC/min up to a maximum reduction temperature of 950 ºC, keeping them at this final temperature for 1 h. The outlet of $H_2$-TPR equipment was connected to a Thermostar GSD301T1 or a PrismaPlus$^{TM}$ mass spectrometer of Pfeiffer Vacuum. The evolution of the mass/charge ratio (m/z) = 18 was used to monitor the formation of $H_2O$ during the $H_2$-TPR process.

UV-Vis specular and diffuse reflectance measurements were carried out using an SHIMADZU 2600i UV–Vis-NIR double-beam spectrophotometer. The spectra in 220 - 1400 nm range were registered in an integrating sphere. Diffuse reflectance spectra were transformed into apparent absorption spectra using the Kubelka–Munk function (*F(R)*). The direct and indirect optical band gap of materials was determined through construction of Tauc plots by plotting *(F(R)hv)$^n$* against *(hv)*, with *n=2 or n=½*, for direct and indirect transitions, respectively. The optical band gap was obtained by extrapolating the linear part of this plot to the energy axis.

Fourier transform infrared (FTIR) spectra of the samples were obtained on a Nicolet 6700 instrument. Samples suspended or diluted in water were added dropwise to a compressed potassium bromide IR-transparent pellet, and then dried to remove all water. FTIR spectra were registered in the range of 4000-500 cm$^{-1}$.

Electron Paramagnetic Resonance (EPR) spectra were measured in a dark room on a Bruker ELEXSYS E500 spectroscope. Samples were prepared by dispersing 3 mg/mL PrO$_x$ in pH 4 acetate buffer. 5,5-dimethyl-1-pyrroline-N-oxide (DMPO) was used as the spin trap for oxygen vacancies while 2,2,6,6-Tetramethylpiperidinooxy (TEMPO) was used for photo-induced electron, which in turn indicates the generation of hole$^+$. A light source from a 300 W xenon lamp (PLS-SXE300, BOFEILAI) was used to generate light at the wavelength range of 320 – 780 nm.

The generation of HO• over PrO$_x$-C sample was monitored by using a fluorescent probe terephthalic acid (TA) which can react with the radicals to form a highly fluorescent product, 2-hydroxy TA (TAOH). H$_2$O$_2$ was added to generate HO• radicals as negative (without PrO$_x$-C) and positive control (with PrO$_x$-C). The fluorescence at 435 nm of TAOH was measured with excitation at 315 nm via a fluorescent spectrometer (FluoroMax-4, Horiba Jobin Yvon, France).

*2.3. Artificial enzyme activities*

The oxidase mimetic activities for the oxidation of colorless TMB, ABTS, DOPA and OPD were studied over three kinds of PrO$_x$ catalysts. The oxidized TMB, ABTS, DOPA and OPD have characteristic UV-vis absorbance peaks at 652, 405, 450 and 450 nm, respectively. All these organic chemicals were bought from Sigma Aldrich. The PrO$_x$ suspension was prepared dispersing powder samples in Milli Q water after washing and centrifuging at 12000 rpm for 10 min three times. All suspensions were sonicated in a 100 W water bath sonicator (Kunshan, KQ-100) for 2 h before each measurement. An extra-strong 500 W ultrasonic processor (VC505, Sonics & Materials, Inc.) was also employed to examine the possible effect of aggregation. Radical scavengers, including TEMPO and p-benzoquinone, isopropanol (hydroxyl HO•  scavenger) and sodium oxalate (hole$^+$ scavenger), were purchased from Sigma Aldrich.

A 4.16 mM TMB stock solution was prepared in dimethyl sulfoxide. Then, a 0.2 mM TMB solution was prepared by diluting the concentrated stock solution with 10 mM pH 4.0 of acetate buffer. Similarly, 0.088 mM ABTS, 0.31 mM DOPA and 4 mM OPD solutions were prepared in 10 mM acetate buffer pH 4.0. The UV-Vis absorption spectra of substrates before and after praseodymia addition were recorded in a SHIMADZU UV-2450 spectrophotometer. To explore and compare the kinetic activities of different samples, praseodymia amount was kept constant at 20 μg/mL with TMB at 0.05, 0.1, 0.15, 0.2, 0.25, 0.3, 0.35 mM to calculate Michaelis-Menten parameter $K_m$ and maximum reaction rate $V_{max}$.

Inhibition effect of L-cysteine was measured in solution of TMB and 45 μg/mL PrO$_x$-NR. Concentration of L-cysteine was varied from 0 to 10 μM. Effects of other amino acids, including 4 μM lysine (Lys), phenylalanine (Phe), arginine (Arg), histidine (His), tyrosine (Tyr), Homocysteine (Hcy), glutathione (GSH) and glycine (Gly), were compared with 4 μM L-cysteine at the same conditions.

*2.4. Fluoride effect and detection*

To compare impacts of several anions, including $F^-$, on the oxidase activity of $PrO_x$-NR, 90 μM TMB and 90 μg/mL $PrO_x$-NR suspension, were added to 5 mM NaF, $Na_2SO_4$, NaCl, $Na_3PO_4$, $NH_4Br$ and $Na_2CO_3$ solutions. Absorbance changes @ 652 nm were measured. For example, the quantitative detection of fluoride in commercial toothpaste (Oral B from supermarket) was measured as follows. Briefly, 1 g toothpaste was vigorously mixed with 32 mL water and filtered through a Millipore Amicon Ultra-15 (3 kDa molecular weight cutoff) ultrafiltration membrane. The filtrate was centrifuged at 15000 rpm for 10 min and the supernatant was collected after centrifugation. Standard addition method [16] was used to analyze fluoride concentration while minimizing the matrix effect. NaF of 0, 10.7, 21.4, 32.2, 42.9, 53.9, and 64.3 μM concentrations were added to 1 mL solution of 90 μg/mL $PrO_x$-NR in pH 4.0 10 mM acetate buffer to form spiked samples. Then 30 μL of collected supernatant was added to the mixture and stirred for 10 min. Finally, 90 μM TMB was added and absorbance at 652 nm was measured.

**3. Results and discussion**

Two types of $PrO_x$ samples, *i.e.* $PrO_x$-NR and $PrO_x$-NP were synthesized and characterized, together with a commercial $PrO_x$-C sample. All of them are dark powders (Fig. 1a), but both $PrO_x$-NR and $PrO_x$-NP samples are more brown-colored than $PrO_x$-C. Their XRD patterns (Fig. 1b) show that all the diffraction peaks can be attributed to a fluorite-like structure, face-centered cubic lattice $Pr_6O_{11}$ (Fm-3m, JCPDS 42-1121).

The diffraction peaks of commercial $PrO_x$-C are much narrower than those of $PrO_x$-NR and $PrO_x$-NP, indicating that the crystallinity of $PrO_x$-C is much better than those of the synthesized $PrO_x$ samples. Using diffraction peak at 28º, the average particle sizes of $PrO_x$-NR, $PrO_x$-NP and $PrO_x$-C were calculated by Scherrer equation to be 7.1, 10.3 and 49.8 nm, respectively.

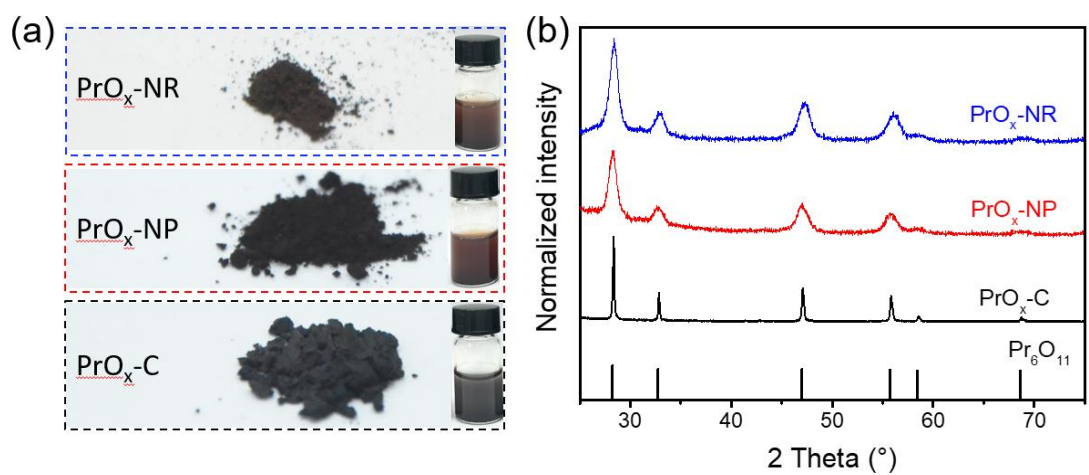

**Fig. 1.** (a) Photographs of three types of $PrO_x$ samples in solid state and suspension in water, including the $PrO_x$-NR synthesized by hydrothermal method, $PrO_x$-NP synthesized by precipitation and commercial $PrO_x$-C. (b) XRD patterns of three $PrO_x$ samples.

Fig. 2 shows SEM images of the three $PrO_x$ samples at low and high magnifications. These samples show different morphologies. $PrO_x$-NR sample prepared by the hydrothermal route presents uniform NR shape. The $PrO_x$ NRs, with diameters in the 20 to 65 nm range and length up to 10 μm, are entangled (Figs. 2a and 2b). $PrO_x$-NP synthesized by precipitation features irregular morphology particles or agglomerates with size less than 5 μm (Fig. 2e). High resolution SEM image (Fig. 2f) of this sample indicates its surfaces of uneven roughness. Commercial $PrO_x$-C also appears to be constituted by agglomeration of large irregularly-shaped particles. $PrO_x$-C particles (Figs. 2i-2j) are not only larger but also show much smoother surfaces than those of PrO$_x$-NP.

TEM and STEM-HAADF images of the three PrO$_x$ samples are also illustrated in Fig. 2. Low magnification TEM image of PrO$_x$-NR (Fig. 2c) and STEM images of PrO$_x$-NP (Fig. 2g) and PrO$_x$-C (Fig. 2k) samples confirm the size and morphology results observed by SEM. The crystalline nature of all the samples is clear appreciated in HRTEM and STEM-HAADF images at the bottom row of Fig. 2. Digital diffraction pattern (DDP) analysis of the selected area of HRTEM image of PrO$_x$-NR (inset in Fig. 2d) shows the major reflections of the fluorite type structure, which is characteristic of higher praseodymium oxide (eg. PrO$_2$). Similar to PrO$_x$-NR catalyst, DDP analysis of selected areas of PrO$_x$-NP (Fig. 2h) and PrO$_x$-C (Fig. 2l) catalysts clearly confirm their fluorite type phase. Interestingly, the DDP of PrO$_x$-C shows the presence of additional spots that it is a clear indication of a superstructure formed by the ordering of oxygen vacancies. A more detailed explanation is included in the supporting information (Fig. S1a). In addition, the PrO$_x$-NP sample is comprised of small crystals in nanometer scale as shown in Figs S1b, which is in good agreement with average particle size of 10.3 nm of PrO$_x$-NP calculated using Scherrer equation.

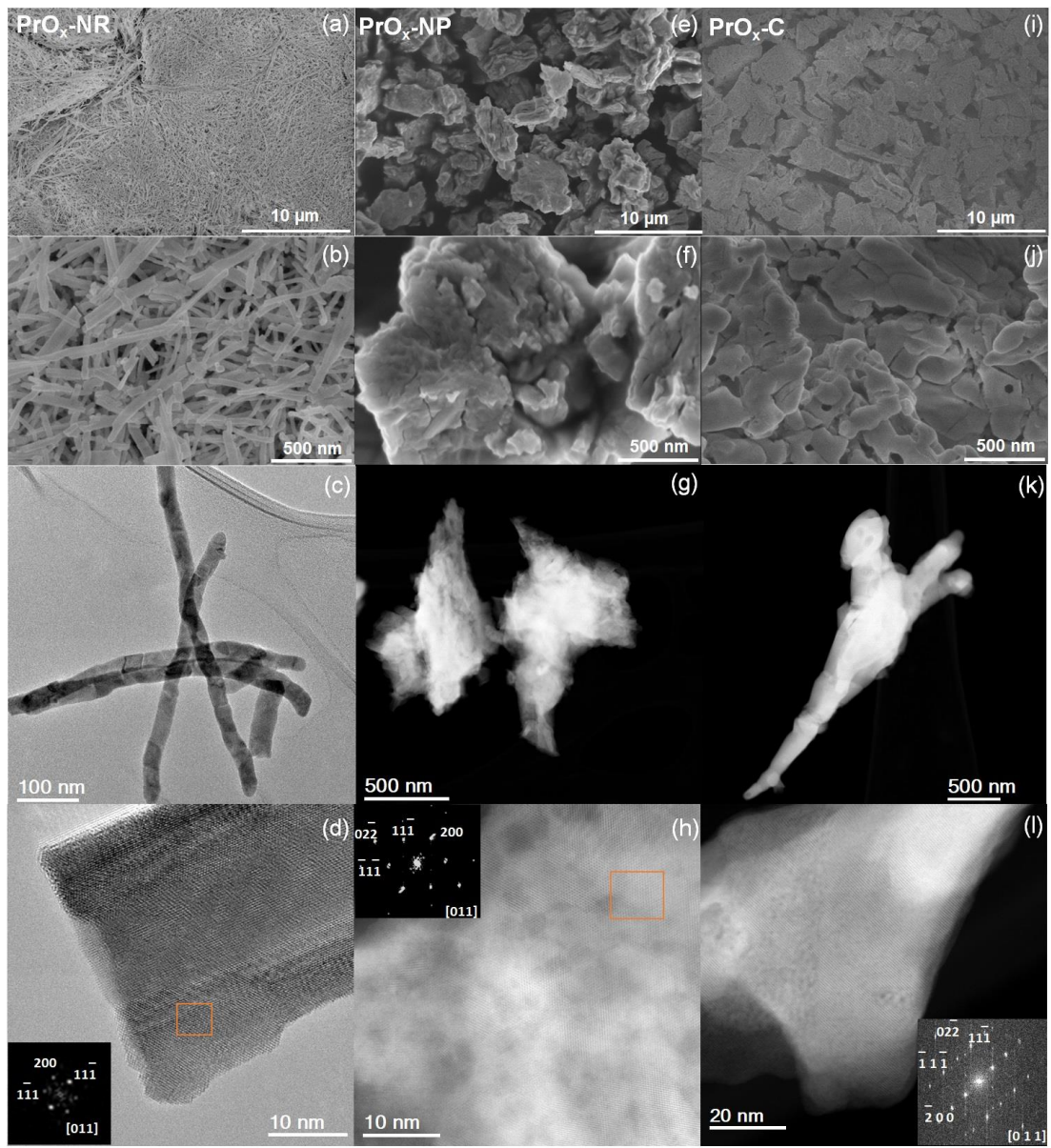

**Fig. 2.** SEM, TEM and STEM-HAADF images of (a-d) $PrO_x$-NR, (e-h) $PrO_x$-NP and (i-l) $PrO_x$-C samples

BET specific surface areas and total pore volumes of three samples were measured by $N_2$ physisorption (Table 1). $PrO_x$-NP possesses the highest surface area, 65 m$^2$/g, followed by $PrO_x$-NR with 27 m$^2$/g and, finally, $PrO_x$-C with just 4 m$^2$/g, a value characteristic of typical bulk materials. The total pore volume data of these samples follow the same trend. These results are in good agreement with the above-mentioned XRD and TEM results. Poor crystallinity and small size of $PrO_x$-NR and $PrO_x$-NP lead to their high BET specific surface areas, while $PrO_x$-C presents bigger particle size, much better crystallinity and much lower surface area.

XPS spectra of three $PrO_x$ samples are shown in Fig. 3 for Pr *3d*, O *1s* and C *1s* core levels. For Pr *3d* core level (Fig. 3a), peak deconvolution was performed by following the analysis reported by Sinev et al. [17]. All spectra could be fitted using 7 peaks, i.e. peaks labelled as *a*, *b* and *c* are for $3d_{5/2}$ component, and *a'*, *b'* and *c'* peaks are their corresponding $3d_{3/2}$ counterparts. An extra peak, existing only in the $3d_{3/2}$ component, and labelled as *t'*, was also included as reported previously [17]. There is no agreement yet on the assignment of peaks to $Pr^{3+}$ or $Pr^{4+}$ oxidation states unequivocally in the literature [17-19]. However, Borchert et al. [18] proposed a method to determine the oxidation degree of Pr by using peaks *a-a'*, which are undoubtedly related with $Pr^{4+}$, and *b-b'*, which appear for both oxidation states. Although they warned that the calculation could lead to a rough approximation, this is one of the most consistent attempts to determine Pr oxidation state using XPS. Here we use this method to calculate $Pr^{3+}$ percentage and results in Table 1 indicate the mixed valence nature of three $PrO_x$ samples with all surface $Pr^{3+}$ percentage at 60 – 70%.

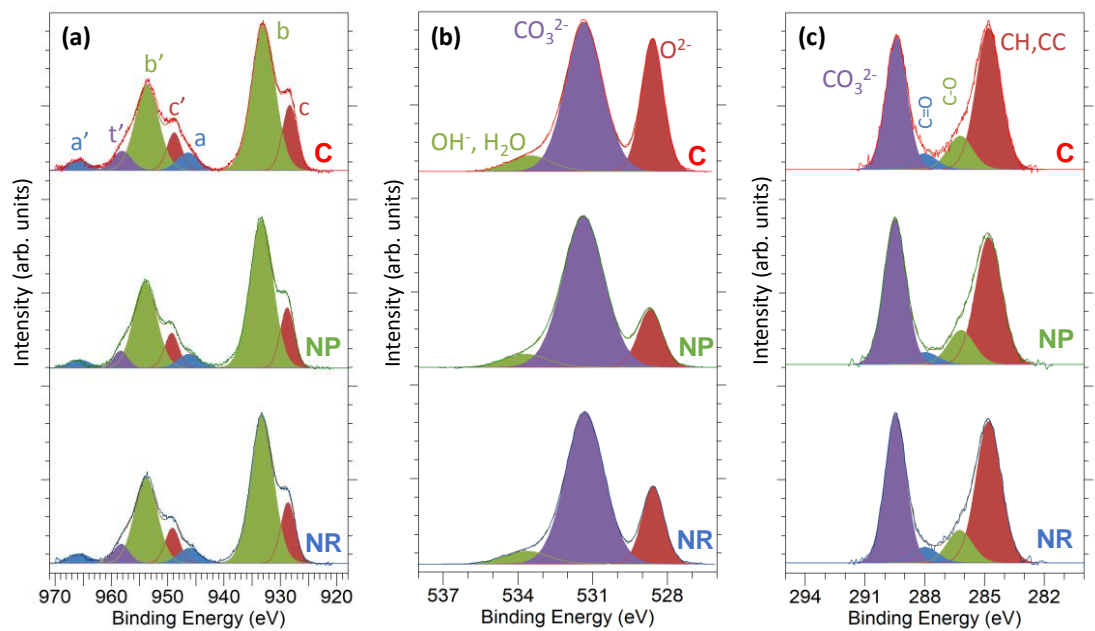

**Fig. 3.** XPS spectra of (a) Pr *3d*, (b) O *1s* and (c) C *1s* of PrO$_x$-NR (blue), PrO$_x$-NP (green) and PrO$_x$-C (red) samples.

Table 1 Physicochemical properties and catalytic activities of PrO$_x$ samples

| Catalyst | BET surface area[a] ($m^2g^{-1}$) | Total pore volume[a] ($cm^3g^{-1}$) | Percentage of $Pr^{3+}$ [b] (%) | Ratio of $O_{carbonate}/O_{lattice}$ [b] | Indirect band gap[c] (eV) | Direct band gap[c] (eV) | $K_m$ [d] (mM) |
|---|---|---|---|---|---|---|---|
| PrO$_x$-NR | 27 | 0.062 | 66 | 3.4 | 1.10 | 2.06 | 0.026 |
| PrO$_x$-NP | 65 | 0.530 | 68 | 4.5 | 0.79 | 2.12 | 0.018 |
| PrO$_x$-C  | 4  | 0.021 | 61 | 1.9 | 0.77 | 2.20 | 0.180 |

[a] Obtained from N$_2$ physisorption

[b] Calculated using XPS data

[c] Determined using Tauc method

[d] Kinetic parameters using TMB as substrate

Further insight about the chemical composition of samples is given by analyzing O *1s* and C *1s* core levels (Fig. 3b and 3c). It is well known that praseodymia reacts readily with atmospheric water and CO$_2$ when it is exposed to air [20, 21]. Analysis of O *1s* and C *1s* core levels confirms these phenomena. O *1s* spectra in Fig. 3b can be fitted using a combination of 3 peaks: one peak around 528.6 eV that could be attributed to lattice oxygen (O$^{2-}$); a second feature, the most intense in all the samples, around 531.3-531.4 eV that can be assigned to carbonates ($CO_3^{2-}$); and a third weak peak at 533.5-533.7 eV, due to the presence of $OH^-$ or water groups. Relative intensity of oxygen from carbonate and lattice after fitting is presented in Table 1. It shows that there could be a correlation between surface area and $O_{carbonate}/O_{lattice}$, i.e. the higher the surface area, the larger the carbonation extent.

C *1s* core level in Fig. 3c also confirms the existence of carbonate on all samples. Peaks at 284.8 eV, 286.2 eV and 288.0 eV correspond to C-C, C-O and C=O, respectively, from adventitious carbon. The dominant peak at 289.4-289.5 eV is due to carbonate. The presence of these carbonated species is strongly related to the high $Pr^{3+}$ content of these samples as determined by XPS. In short, the higher the surface area of the $PrO_x$ samples, the higher the carbonate content, and thus the higher the reduction degree of praseodymium.

Fig. 4 shows UV-visible diffuse reflectance spectra of $PrO_x$ catalysts. Absorbance spectra of $PrO_x$-NP and $PrO_x$-C samples are very similar in 220-1400 nm wavelength range. In comparison, $PrO_x$-NR exhibits a quite strong absorbance in the visible range between 220 to 640 nm, in agreement with previous reports [12]. This is probably why $PrO_x$-NR appears more brown-colored than the other two samples (Fig. 1a). As shown in XRD results, the main phase of three $PrO_x$ catalysts is $Pr_6O_{11}$, which contains both $Pr^{3+}$ and $Pr^{4+}$ oxidation states. The contribution of $Pr^{3+}$ and $Pr^{4+}$ absorbance in UV-visible range is still in debate. Kang et al. reported that the presence of oxidation state of $Pr^{4+}$ may be the origin of the strong/broad absorption in the visible region by charge transfer ($O^{2-} \rightarrow Pr^{4+}$) absorption [12]. On the other hand, previous studies also reported $Pr^{4+}$ does not absorb in the UV-vis region [22]. In other words, the absorbance in the 430 – 610 range can be mainly attributed to $Pr^{3+}$. Meanwhile, $Pr^{3+}$ absorption bands between 430 – 490 nm due to the $^3H_4$ ground state to $^3P_J$ transition and that the 560 – 610 nm band originated from a $^3H_4$ to $^1D_2$ transition [22]. It should be highlighted that there is no big difference between $Pr^{3+}$ percentages (Table 1) of $PrO_x$-NR and $PrO_x$-NP catalysts calculated using XPS results, and $Pr^{3+}$ percentage of $PrO_x$-C is slightly lower than $PrO_x$ with NR and NP morphologies. The reason of UV-visible strong absorption of $PrO_x$-NR different with $PrO_x$-NP and $PrO_x$-C is not clear, which needs further study in future.

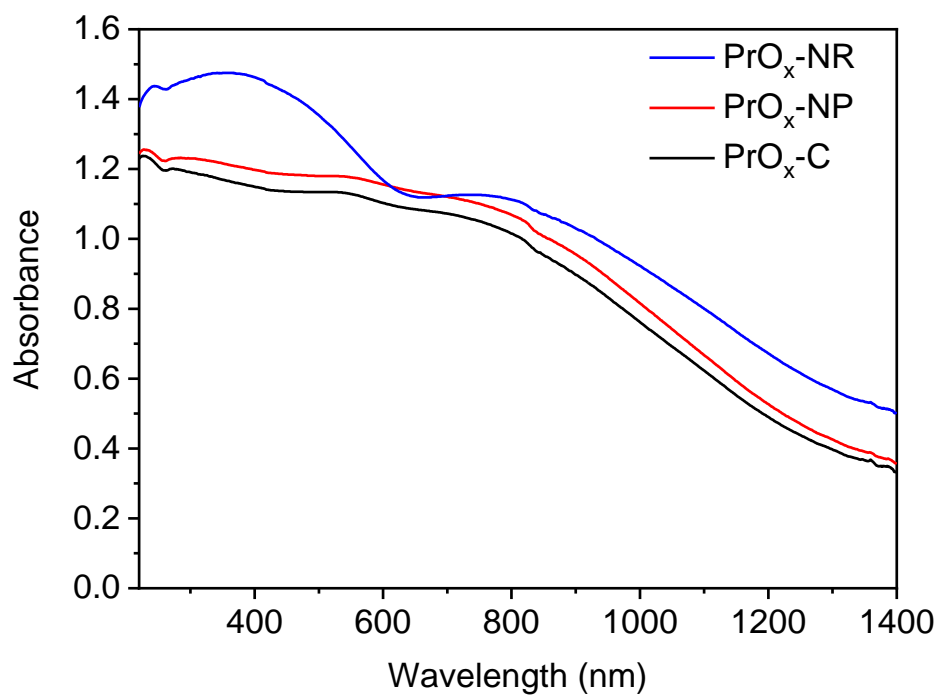

**Fig. 4.** UV-Visible diffuse reflectance spectra of $PrO_x$ samples.

Using the UV-Vis diffuse reflectance results, indirect and direct band gap energies of $PrO_x$ samples can be determined via the Kubelka-Munk function. As shown in Table 1, there is no significant difference in direct band gap values (2.06 – 2.20 eV) for three PrO$_x$ samples, in accordance with previous results in the literature [10, 23]. However, some researchers have reported higher values for direct band gap of PrO$_x$, from 3.15 to 4.45 eV [24]. The origin of such difference in optical properties and band gap energies could stem from the synthesis methods of PrO$_x$. The band gap reduction may be attributed to both the uplift of valence band (due to surface disorder) and lowering of conducting band (due to oxygen vacancies and more defect centers) [25].

Fig. 5 displays the formation of H$_2$O over all PrO$_x$ samples during H$_2$-TPR process. The redox behavior of PrO$_x$-NR is better than those of the other two samples. Its dominant reduction peak appears at 480 °C, at lower temperature than in PrO$_x$-NP (502 °C) and PrO$_x$-C (509 °C). The reason might be due to its smaller average crystalline size (7.1 nm) than those of other two samples (10.3 and 49.8 nm) according to calculation using Scherrer equation. The main reduction peak corresponds to the reduction of Pr$^{4+}$ in bulk of PrO$_x$ samples. TPR profiles of PrO$_x$-NP and PrO$_x$-C catalysts are similar in that both have a minor reduction peak at 310 or 332 °C, which can be assigned to the reduction of surface Pr$^{4+}$ and the active surface oxygen species react with hydrogen to form H$_2$O [12, 26]. The reduction peak of PrO$_x$-NR at low temperature is merged into an asymmetric peak at 480 °C. This result is in good accordance with previous TPR data of PrO$_x$-NR and bulk PrO$_x$ in the literature [26]. The final product of H$_2$-TPR of three PrO$_x$ samples is expected to be Pr$_2$O$_3$, according to literature [8, 21].

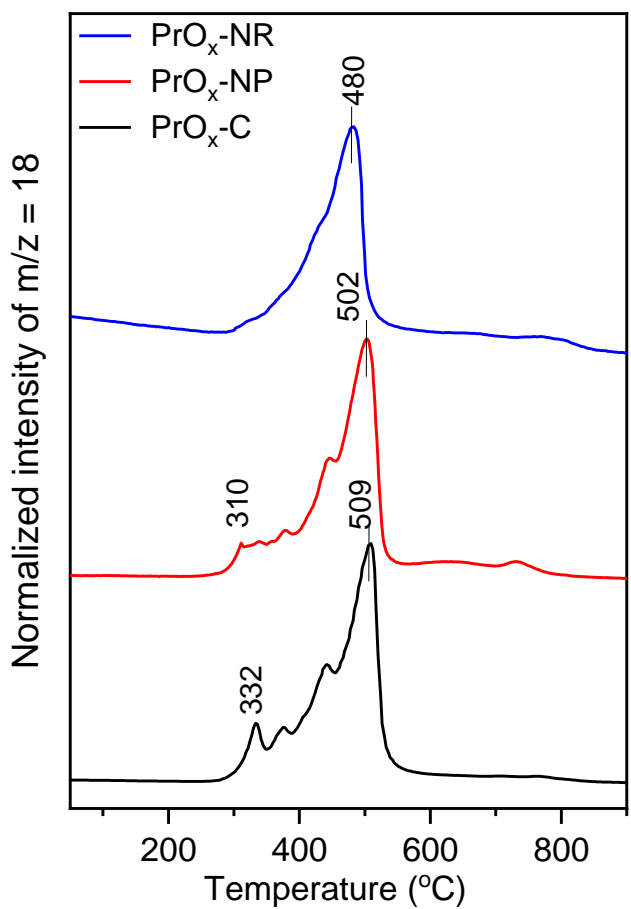

**Fig. 5.** H$_2$O evolution (m/z=18) during H$_2$-TPR of PrO$_x$-NR, PrO$_x$-NP and PrO$_x$-C catalysts.

The multi-functional oxidase-like activity of PrO$_x$ was studied by their effects in oxidizing conventional substrates including TMB, ABTS, OPD and DOPA. TMB, ABTS and OPD are chromogenic substrates with oxidable amino groups, while DOPA represents substrates with polyphenol groups. Once PrO$_x$ samples were added to substrates in pH 4.0 acetate buffers, individual characteristic colors quickly appeared (Fig. 6a), suggesting the occurrence of oxidation of substrates. The corresponding UV-vis absorbance spectra changes of all substrates are illustrated in Fig. S2. Similar to other previously reported oxidase-like nanomaterials, the catalytic activities of PrO$_x$ catalysts are pH-dependent and pH 4.0 is the optimum condition for all substrates (Fig.

6b). In comparison, TMB and DOPA are more sensitive to pH changes than ABTS, probably because they are positively charged while ABTS is negatively charged, inducing different adsorption and catalytic behaviors of nanozymes as previously reported [27].

Kinetics of three $PrO_x$ samples (Fig. S3) can be calculated by the Michaelis-Menten equation as shown in Method in Supporting information using TMB as a substrate and their kinetic parameters $K_m$ are listed in Table 1. Lower $K_m$ value means higher oxidase activity of nanozymes. These results indicate that the oxidase-like activities follow the order of $PrO_x$-NP > $PrO_x$-NR > $PrO_x$-C, exactly the same order of their BET surface areas. More importantly, $K_m$ values of both $PrO_x$-NR (0.026 mM) and $PrO_x$-NP (0.018 mM) are an order of magnitude lower than that of $PrO_x$-C (0.18 mM), and significantly lower than most of the reported oxidase-like nanozymes. Their maximum reaction rates $V_{max}$ also show same tendency, 0.031, 0.029 and 0.0013 mM/min, respectively. Table S1 compares $K_m$ and $V_{max}$ values from this study with those of the so-far-reported nanozymes in TMB oxidation since 2007, including high activity of $Mn_2O_3$ nanoparticles [4]. The comparison suggests that $PrO_x$-NP and $PrO_x$-NR exhibit relatively higher activities than those nanozymes of similar levels of $V_{max}$ (Table S1).

Fig. 6c compares steady-state catalytic activities of three $PrO_x$ catalysts, as well as those of three compounds of praseodymium, $Pr_2O_3$, $Pr_2(CO_3)_3$, and $Pr(OH)_3$. There is almost no oxidase activity of TMB oxidation over these three Pr-containing compounds with solely Pr oxidation state of +3. At the beginning of reaction (< 1 min),

PrO$_x$-NP performs faster than PrO$_x$-NR, but over longer time (> 2 min), PrO$_x$-NR is better than PrO$_x$-NP. This is probably because the aggregated state of PrO$_x$-NP partly inhibits the following attachment of substrates, which was further examined with the help of extra-strong ultrasonicator in Fig. S4. However, PrO$_x$-NR is much better dispersed in solution; therefore, it exhibits better long-term catalytic activity. Furthermore, similar $K_m$ and $V_{max}$ values of PrO$_x$-NR and PrO$_x$-NP means that TMB oxidation occurs with similar kinetic rate, i.e. similar amount of TMB is oxidized over both samples. Often the catalytic activities of nanomaterials can be compared as activity per unit surface area [28]. In this study, TMB oxidation takes place on the surface of PrO$_x$ samples, considering that the BET surface area of PrO$_x$-NR is less than half of PrO$_x$-NP, it means that the activity per unit area on PrO$_x$-NR surface is higher than that of PrO$_x$-NP.

As mentioned before, the Pr$^{3+}$/Pr$^{4+}$ ratio of PrO$_x$ can change easily with storage conditions so that it bears a risk of phase change upon exposure to air after long time storage [21]. Fig. S5a shows XRD patterns of three PrO$_x$ catalysts stored in air for years. Pr(OH)$_3$ phase can be observed on all PrO$_x$ samples. However, PrO$_x$-NR stored for 9 years and PrO$_x$-NP stored for 3 years contain less Pr(OH)$_3$ phase and more PrO$_2$ and Pr$_6$O$_{11}$ phases than PrO$_x$-C stored for 5 years. Since calcination of Pr(OH)s is a step of synthesis of PrO$_x$, it is reasonable to recalcine PrO$_x$ samples exposed in air for a long time at 500 °C to obtain pure PrO$_x$ phase again. This hypothesis is confirmed, which calcination at 500 °C leads to the formation of pure Pr$_6$O$_{11}$ phase again as shown in Fig. S5a. Fig. S5b shows the TMB oxidation reactions by PrO$_x$-NP and PrO$_x$-C after three-year storage. Both samples were recalcined at 500 °C for 4 h to regenerate $Pr_6O_{11}$ and effective TMB oxidation is observed, suggesting the good stability of $PrO_x$ samples.

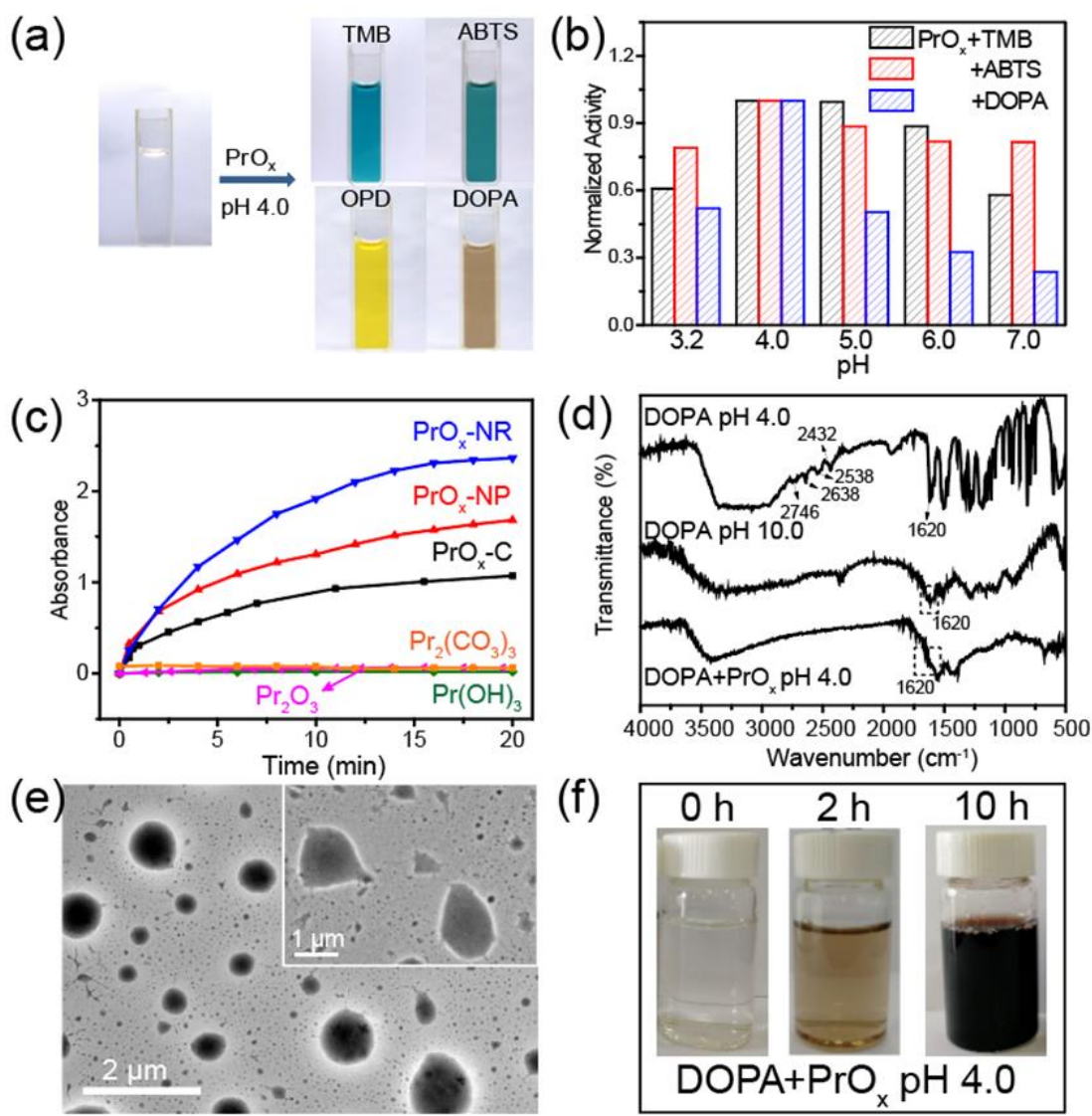

**Fig. 6.** (a) Colorless TMB, ABTS, OPD and DOPA substrates can be catalytically oxidized by $PrO_x$ to generate characteristic colors at pH 4.0. (b) pH-dependence of oxidase-like activities of $PrO_x$-NR samples. Normalized activity of each reaction was obtained by normalizing absorbance of oxidized TMB at 652 nm, oxidized ABTS at 405 nm and oxidized DOPA at 450 nm by absorbance of un-oxidized substrates, respectively. (c) Absorbance changes at 652 nm of 0.2 mM TMB in pH 4.0 acetate buffer when 92 μg/mL catalysts ($PrO_x$, $Pr_2O_3$, $Pr_2(CO_3)_3$, and $Pr(OH)_3$) were added. (d) FTIR spectra of 1 M DOPA at conditions of pH 4.0, pH 10.0 and pH 4.0 with 1.1 mg/mL $PrO_x$-NR. (e) Optical microscopic image of DOPA solution after addition of $PrO_x$-NR in pH 4.0 solution, forming dark polydopamine spheres. Inset is the image of DOPA solution without $PrO_x$-NR. (f) Photographs of $PrO_x$-NR oxidizing and polymerizing dopamine in acidic condition.

Since PrO$_x$-NR nanozyme presents the highest steady-state activity and similar $K_m$ value to PrO$_x$-NP, further studies to understand mechanism and influences of amino acids and anions focus on this catalyst. The activities of PrO$_x$-NR catalysts are so high that they can even enable the polymerization of concentrated DOPA to polydopamine (PDA) in acidic solution. FTIR spectra of DOPA at pH 4.0 and 10.0 and pH 4.0 with PrO$_x$-NR conditions are compared in Fig. 6d. The characteristic bands at 2746, 2638, 2538 and 2432 cm$^{-1}$ of DOPA disappear at pH of 10.0 and at pH of 4.0 with PrO$_x$-NR, because the polymerization of DOPA changes NH$_2$ to secondary amine [29]. Meanwhile, the width of 1620 cm$^{-1}$ peak increases after polymerization at pH of 10.0 and at pH of 4.0 with addition of PrO$_x$-NR, probably due to C=C stretching vibration of indole structure formed by intramolecular cyclization reactions [30]. Further, results in Fig. 6e and 6f also demonstrate the gradual oxidation and induced-polymerization of DOPA.

It is well known that DOPA can be oxidized and self-polymerized to PDA in basic (pH >8) media, but not in neutral and acidic media [31-34]. PDA is a wet bio-adhesive that has unique photo-physical properties for a wide range of biomedical applications. To oxidize and polymerize PDA in acidic condition usually involves unconventional tools such as UV-illumination [31, 32], plasma treatment [33], high temperature (over 120 °C for 16 h) [34]. To the best of our knowledge, this is the first time that oxidative self-polymerization of PDA is achieved in an acidic environment with solely the help of inorganic nanoparticles as catalysts. It is worth noting that an employment of PrO$_x$-NR as oxidase to synthesize PDA may also be beneficial to widen the pH range of application of PDA.

The key reasons for such high catalytic performance of $PrO_x$ may be attributed to the important intermediate holes$^+$, which is remarkably different from other nanozymes. So far most reported oxidase-like nanozymes, such as $CeO_2$, $MnO_2$, iron-based materials etc. [2], interact with molecular oxygen (or other oxidizing reagents) to generate intermediate superoxide radical anion ($O_2^{•-}$) as the key oxidants in the oxidation process. In order to elucidate the mechanism of oxidase-like property of $PrO_x$-NR, several scavengers corresponding to different reactive oxygen species and charge carriers were applied to $PrO_x$-TMB system. For example, p-benzoquinone [35] in Fig. 7a and TEMPO [36] in Fig. 7b were used to capture $O_2^{•-}$, while isopropanol [37] is a hydroxyl (HO$^•$) scavenger and sodium oxalate [38] was added to eliminate hole$^+$ [39]. Results show that only sodium oxalate (hole$^+$ inhibitor) can significantly affect the oxidase-like activity and the inhibition level is highly dependent on the amount of eliminator. On the contrary, the other three inhibitors show no influence within the concentration range studied in this work. These results imply that, in $PrO_x$-catalyzed oxidation, the key intermediate is hole$^+$, not the conventional radicals like $O_2^{•-}$ or HO$^•$. To the best of our knowledge, this is the first time that hole$^+$ was reported for the activity of nanozyme. This feature of $PrO_x$ is also in clear contrast to its affiliated oxide $CeO_2$, which is well known for generating $O_2^{•-}$. Such hole$^+$-induced oxidation is also dependent on pH values (Fig. 6b), probably because that the surface hydroxyls can react with hole$^+$ to form hydroxyl radicals and thus the total oxidative capacity decreases at higher pH conditions.

The generation of holes$^+$ was further tested by EPR spectra. It was known that photoexcitation may lead to the concomitant generation of electron/hole$^+$ pairs [40]. The identification of photoinduced electrons by EPR is a clear indication of the presence of hole$^+$. The spin label TEMPO does not react with oxidative intermediates such as $O_2^{\bullet-}$ or $HO^{\bullet}$ [40]. Therefore, it is good capture probe for photoinduced electrons, as well as indirect indication of hole$^+$. Fig. 7c shows that TEMPO signal (g value ~ 2.000) intensity reduced 14% after 5 min light illumination, and almost disappeared after 20 min illumination, suggesting the generation of electrons and hence holes$^+$. Interestingly, EPR measurements also demonstrate the involvement of oxygen vacancies, $VO^{\bullet\bullet}$, in $PrO_x$-catalyzed oxidation. Using DMPO as spin label, either using ABTS or TMB substrate, Fig. 7d shows clear reduction of $VO^{\bullet\bullet}$ (g value ~ 2.003) signals when sodium oxalate, the hole$^+$ inhibitor, were added to the mixture. The oxidation of TMB over $PrO_x$-C in dark and under day light (Fig. S7) have been studied. Results showed that the enzymatic activity for oxidation of TMB is lower in the dark compared with under day light, suggesting the possible influence of photo-induced hole$^+$.

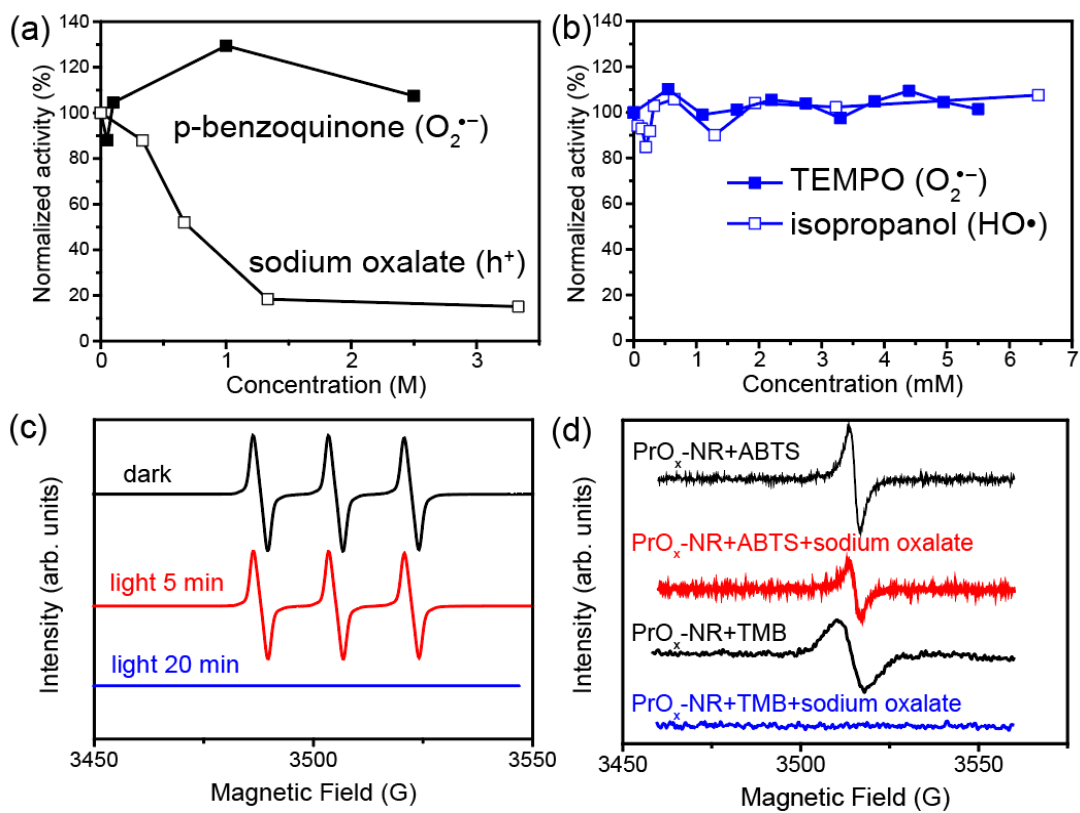

**Fig. 7.** (a) Normalized catalytic activity of PrO$_x$-NR oxidizing TMB after adding p-benzoquinone (O$_2^{\bullet-}$ scavenger) and sodium oxalate (hole$^+$ scavenger). (b) Effect of isopropanol (HO$^\bullet$ scavenger) and TEMPO (O$_2^{\bullet-}$ scavenger). (c) EPR spectra of spin label TEMPO for electrons generated by 3 mg/mL PrO$_x$-NR pH 4.0 in dark and after light illumination of 5 and 20 min. (d) EPR spectra of spin label DMPO for 3 mg/mL PrO$_x$-NR with 1 mM ABTS and 0.1 mM TMB before and after addition of 20 mM sodium oxalate.

Taken together, the above-mentioned results suggest that the catalytic oxidation, or the oxidase-like activity, of PrO$_x$ is closely related to hole$^+$ and VO$^{\bullet\bullet}$. Energy level diagram (Fig. 8a) of PrO$_x$-NR demonstrates that the distance between O-2p Valence Band (VB) (2.81 eV) and Pr 4f Conduction Band (CB) is 2.06 eV, and VO$^{\bullet\bullet}$ is located between these two band gaps. The level of VO$^{\bullet\bullet}$ can be estimated from the photon energy profile (Fig. 8b) based on UV-Vis diffuse reflectance data in Fig. 4. Each absorption band can be associated with an electron transition between two energy states (Fig. 8a). For example, the absorption bands at 1.29 and 1.55 eV could be assigned to the charge transfer transition from oxygen vacancies to Pr 4f CB. Fig. 8c shows VB

XPS of PrO$_x$-NR in comparison with a 100% reduced praseodymium oxide [3]. Their O-2p VB maximum are similar at around 2.81 eV.

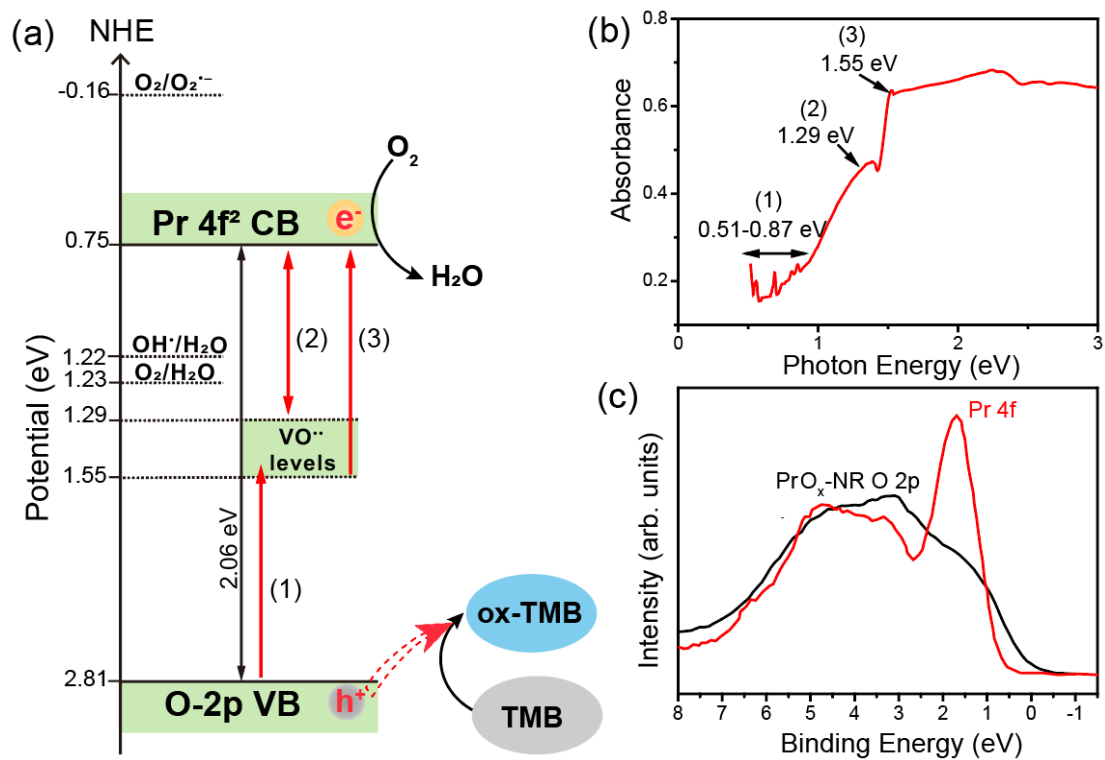

**Fig. 8.** (a) Energy level diagram and excitation process of PrO$_x$-NR. (b) Photon energy spectrum of PrO$_x$-NR. (c) Valence band XPS data for a 100% reduced praseodymium oxide and PrO$_x$-NR.

The mechanism for PrO$_x$-catalyzed oxidation is proposed to be initiated by electrons transferring from O-2p VB to the VO$^{\bullet\bullet}$ band, leaving a hole$^+$ available to oxidize substrates like TMB and DOPA (Fig. 8a). Meanwhile, the electron can be further transferred to the conduction band and O$_2$, which is eventually reduced to H$_2$O. The formation of O$_2^{\bullet-}$ from O$_2$ requires the potential of electron to be less than -0.16 eV [40], while the standard reduction potential for the O$_2$/H$_2$O couple is 1.23 V (Fig. 8a) [41]. It is likely that the electron can reduce O$_2$ to H$_2$O without generating O$_2^{\bullet-}$. On the other side, although hole$^+$ has a greater potential (2.81 eV) than HO$^{\bullet}$ (2.2 eV), the inhibition measurement (Fig. 7b) shows no presence of HO$^{\bullet}$ radicals. Therefore, we propose that hole$^+$ may probably oxidize the substrate directly. Moreover, the presence of VO$^{\bullet\bullet}$ and high oxygen mobility of PrO$_x$ are able to facilitate the electron/hole$^+$ separation, which leads to the high oxidizing capability of PrO$_x$ catalyst.

Despite the proposed mechanism above, it is worth noting that the oxidase-like property of PrO$_x$ is probably not only the result of a single factor. The hole$^+$ in PrO$_x$ crystalline lattice can be also generated by various oxidative species (HO•, O$_2^{\bullet-}$, etc), directly or indirectly participating in the oxidation (depicted as dotted arrow in Fig. 8a). For example, although the addition of HO• inhibitor (isopropanol) shows no impact on the oxidase-like activity of PrO$_x$ (Fig. 7b) at increasing concentration, fluorescent experiments using HO• probe terephthalic acid (TA) indicate the presence of a small amount of HO• radical compared with negative control without PrO$_x$-C and positive control with PrO$_x$-C experiments (Fig. S8). The TA aqueous solution and a mixture of TA and H$_2$O$_2$ solution do not show a fluorescent peak at 435 nm, which is a characteristic peak of 2-hydroxy TA (TAOH) generated from reaction between HO• radicals and TA. When PrO$_x$-C was mixed with TA, there is a small fluorescent peak at 435 nm. The addition of H$_2$O$_2$ to the mixture of PrO$_x$-C and TA generates more HO• radicals, thus a stronger peak appears at 435 nm. All these results suggest that the existence of multiple routes for generation of oxidative species, such as the interaction of OH$^-$ with hole$^+$ to form HO• (OH$^-$ + h$^+$→HO•). The origin of oxidase-like activity of PrO$_x$ in the crystalline lattice may be eventually related to the reversible exchange between Pr$^{3+}$ and Pr$^{4+}$ which is associated with the formation and migration of oxygen vacancies. Pr$_6$O$_{11}$ has the highest oxygen mobility among all rare earth oxides [7], partly because of its inherent structure of naturally "doped" $Pr^{3+}$ in $Pr^{4+}$ stoichiometry. The oxygen defects or vacancies, particularly abundant in $PrO_x$-NR of this study, lead to its excellent oxygen storage capacity, which is fundamental for $Pr_6O_{11}$ in previous catalytic studies [9].

$PrO_x$ as oxidase is sensitive to reducing agents, such as L-cysteine. Absorbance of oxidized TMB decreases with addition of L-cysteine (Fig. 9a). In fact, absorbance change @652 nm decreases linearly with L-cysteine concentration within the $2\times10^{-7}$ ~ $1\times10^{-5}$ M range, with a limit of detection (LOD) of $2.3\times10^{-9}$ M (Fig. 9b). In comparison with other nanomaterials employed in various L-cysteine sensors (Table S2) [42-44], $PrO_x$-TMB platform appears to be a good colorimetric sensing probe for cysteine and similarly structured Hcy in blood or serum samples, with negligible influence from salt or amino acids of similar structures (Fig. 9c).

Oxidase activity of $PrO_x$-NR is also very sensitive to the fluoride anion ($F^-$), a high dose of which is toxic to human health and affects the aqueous conditions in the environment [45]. As shown in Fig. 9d, $F^-$ addition inhibits TMB oxidation over $PrO_x$-NR and the inhibition level is in linear relationship to $F^-$ concentration (Fig. 9e). Meanwhile, other halides ($Cl^-$ and $Br^-$) and anions ($SO_4^{2-}$, $PO_4^{3-}$ and $CO_3^{2-}$) do not induce such dramatic effect (Fig. 9f). This high selectivity for $F^-$ ions of $PrO_x$-TMB system can be used as a sensing platform for the excessively harmful $F^-$ anions which could cause dental mottling and skeletal manifestation [46]. For example, this method can be employed to measure the fluoride content in a supermarket toothpaste (Oral-B as in Fig. 9d). To minimize the matrix effects, standard addition method was used (Fig.

S6) [16]. Accordingly, the measured F⁻ content is 0.14%, which is very close to the labelled value (0.15%) on the toothpaste, suggesting good accuracy and applicability of this method in ordinary F⁻ detection.

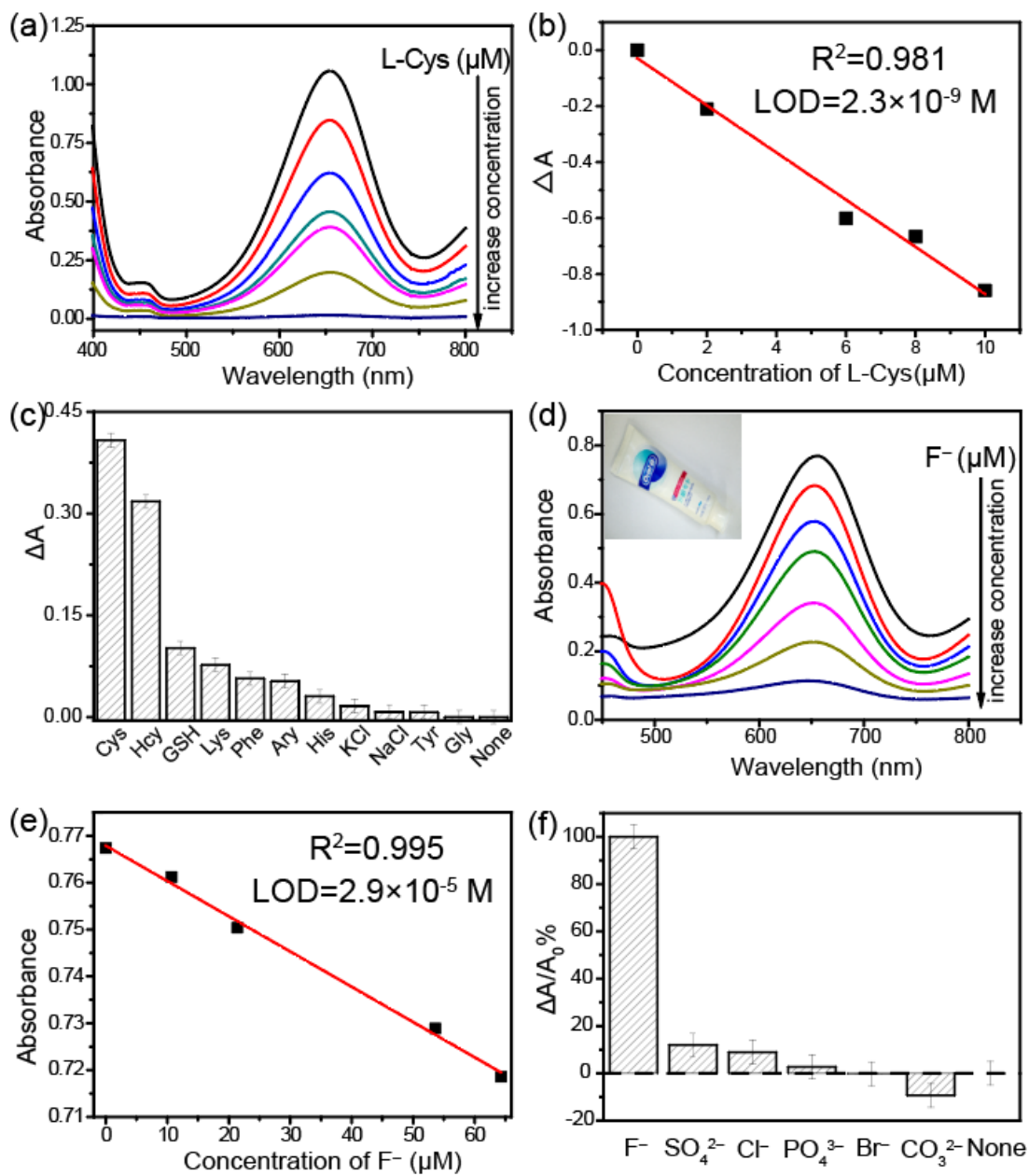

**Fig. 9.** (a) UV-Vis spectra of 0.2 mM TMB oxidized by 0.02 mg/mL PrO$_x$-NR in presence of L-cysteine (0 - 10 μM). (b) Correlation between A$_{652\ nm}$ with L-cysteine concentration with LOD of 2.3×10$^{-9}$ M and R$^2$ = 0.981. (c) Selectivity of L-cysteine detection in comparison with salts and other amino acids. (d) UV-Vis spectra of 0.09 mM TMB oxidized by 0.1 mg/mL PrO$_x$-NR at different F⁻ concentration. Inset is picture of an Oral-B toothpaste containing 0.15%

fluoride. (e) Relationship between $A_{652\ nm}$ with $F^-$ concentration. (f) Selectivity of $F^-$ detection in comparison with other anions.

Another phenomenon worth noting is that the fluoride-induced inhibition on $PrO_x$ is in contrast with the enhancement effect of fluoride on the neighbor oxide $CeO_2$ [47]. Based on previous EPR and Raman studies [48, 49], $F^-$ anions are adsorbed on surface of $CeO_2$, increasing the concentration of $VO^{\bullet\bullet}$, but does not affect the amount of generated $O_2^{\bullet-}$ or $HO^{\bullet}$ in function of mimicking oxidase. However, it can accelerate the oxidation process by replacing the oxidation product quickly on $CeO_2$ surface. Oxidase-like $PrO_x$ also functions through $VO^{\bullet\bullet}$, but its interaction with $F^-$ is different. Fig. 10 shows XPS spectra of $PrO_x$-C before and after mixing with $F^-$. Opposite to $CeO_2$, whose $Ce^{3+}$ concentration increased significantly (from 38% to 45%) after $F^-$ immersion [48, 49], the percentage of $Pr^{3+}$ on $PrO_x$ decreases from 61% to 56%, while the ratio of $O_{carbonate}/O_{lattice}$ decreases from 1.9 to 1.45. Adsorption of F element is also confirmed on $PrO_x$ surface because a new F $1s$ peak at 684 eV appears in XPS of $F^-$-treated $PrO_x$ (Fig. S9) [50]. $F^-$ was reported to bind to hydroxyl groups on surface of metal oxides including Mg-Al [50]. As a strong electron withdrawing ligand, the addition of $F^-$ anions leads to oxidation of some $Pr^{3+}$ to $Pr^{4+}$ in order to keep the charge balance in $PrO_x$. Thus, the active sites $Pr^{3+}$ on $PrO_x$ surface decrease and fluoride-induced inhibition occurs. Secondly, $F^-$ may interrupt the electron transfer pathway and reduce the formation of surface $hole^+$, which also causes the inhibition of $PrO_x$ activity. These are probably the reasons why $F^-$ has very opposite effect on the oxidase-like activity of $PrO_x$ to $CeO_2$.

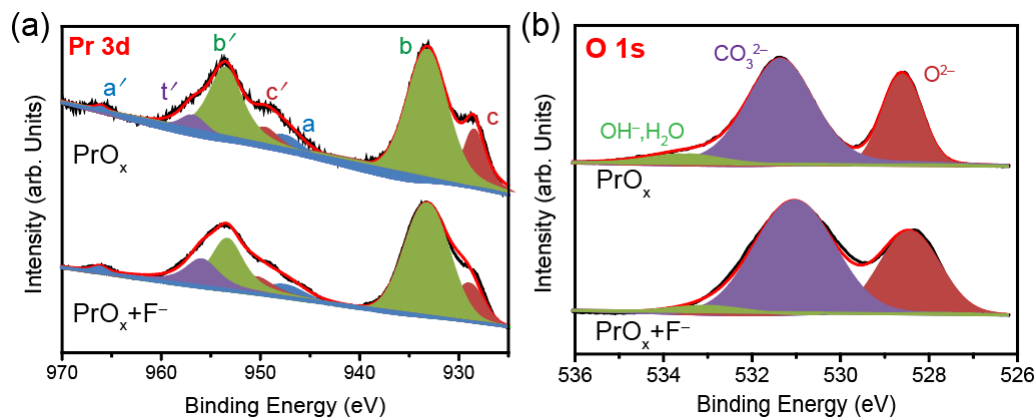

**Fig. 10.** XPS spectra of (a) Pr *3d* and (b) O *1s* of $PrO_x$-C before and after treating with 10 mM $F^-$ solution.

Fig. 11 illustrates schematically the possible mechanism in the multi-functional oxidase-like activity of $PrO_x$. Praseodymium cations can exist with +3 and +4 valence states in $PrO_x$ samples. $PrO_x$ has the highest oxygen mobility because it has a larger number of possible phases for the oxides, so that the variety of stable phases enable fast changes in the oxidation states of praseodymium. $PrO_x$ demonstrates high multi-oxidase activities which can be specifically tuned by its morphology, size, redox property, $Pr^{3+}/Pr^{4+}$ ratio and reducing inhibitor such as L-cysteine. Due to the generated hole$^+$, oxygen vacancies and high oxygen mobility, $PrO_x$-NR is able to initiate the oxidation and polymerization of DOPA in acidic condition. Moreover, adsorption of $F^-$ can inhibit the activity of $PrO_x$-NR, which conversely enables the linear detection of $F^-$. This effect makes $PrO_x$ a good probe for $F^-$ sensing, but with completely different mechanism to $CeO_2$ which detects $F^-$ by the boosted activity. In an environment of weak target presence, it is usually easier to detect reduction of a strong signal (signal-off) than enhancement of a weak signal (signal-on) with possibly high background noise. From this point of view, for $F^-$ detection normally at low concentrations in most testing environments, $PrO_x$-NR is probably better than $CeO_2$ nanomaterials as sensors.

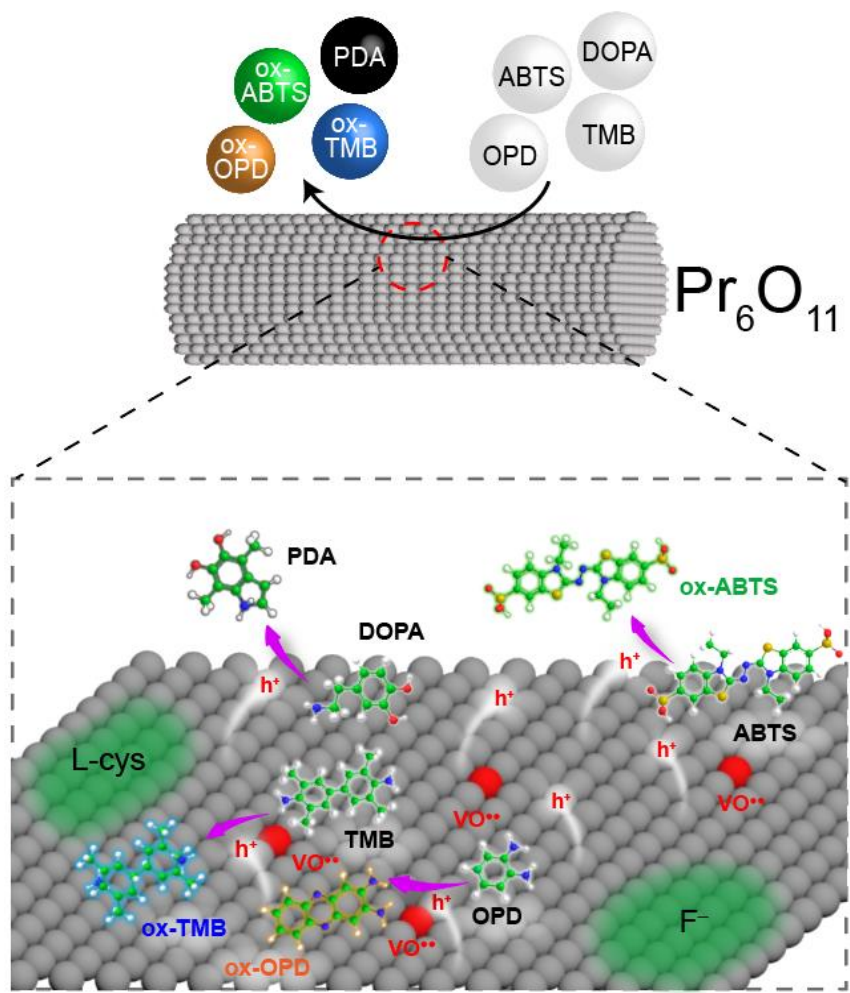

**Fig. 11.** Schematic illustration of how PrO$_x$-NR catalyzes the oxidation of several substrates including TMB, ABTS, OPD and DOPA. The oxidizing activity is high enough to oxidize and polymerize DOPA into PDA in acidic conditions. The key intermediates and reactive species are hole$^+$ and oxygen vacancies VO$^{\bullet\bullet}$, instead of O$_2^{\bullet-}$ or HO$^\bullet$, in the oxidation process.

### 4. Conclusions

PrO$_x$ is firstly reported to exhibit multi-functional oxidase-like activities with significantly high kinetic performance. Three types of PrO$_x$ samples with different morphologies and surface areas are compared in their catalytic activities of oxidizing a series of amino and polyphenol substrates, such as TMB, ABTS, OPD and DOPA. The polymerization of DOPA can be triggered by PrO$_x$-NR under acidic condition to produce polydopamine, a highly useful wet bioadhesive. Different from previously reported oxidase-like nanozymes, hole$^+$ and oxygen vacancies may play important roles in the oxidation process by PrO$_x$. Adsorption of F$^-$ on PrO$_x$ inhibits its activity, in clear contrast to CeO$_2$, probably because of the increased concentration of Pr$^{4+}$ on surface and reduced oxygen accessibility by the F$^-$ blockage. These results not only shed some light on the discovery of REO artificial enzymes with new mechanism, but also imply a new sensor for F$^-$ detection in fields of contamination and overdose.

**CRediT authorship contribution statement**

**Lei Jiang:** Conceptualization, Investigation, Validation, Data curation, Writing – original draft, Methodology, Formal analysis, Writing – review & editing. **Yaning Han:** Data curation, Formal analysis. **Susana Fernández-García:** Investigation, Review & editing. **Miguel Tinoco:** Investigation, Review & editing. **Zhuang Li:** Investigation. **Pengli Nan:** Investigation. **Jingtao Sun:** Investigation. **Juan J. Delgado:** Investigation, Methodology, Review & editing, Funding acquisition, Project administration. **Huiyan Pan:** Investigation. **Ginesa Blanco:** Investigation, Review & editing. **Javier Martínez-López:** Investigation. **Ana B. Hungría:** Supervision, Review & editing. **Jose J. Calvino:** Supervision, Review & editing. **Xiaowei Chen:** Conceptualization, Methodology, Investigation, Funding acquisition, Project administration, Resources, Supervision, Data curation, Writing – review & editing.

**Declaration of Competing Interest**

The authors declare that they have no known competing financial interests or personal relationships that could have appeared to influence the work reported in this paper.

**Acknowledgements**

This work has been supported by the Ministry of Science, Innovation and Universities of Spain with reference number of PID2020-113809RB-C33 and Junta de Andalucía (Spain) with reference number of PY18-2727. This work has been co-financed by the 2014-2020 ERDF Operational Program and by the Department of Economy, Knowledge, Business and University of the Regional Government of Andalusia with project reference FEDER-UCA18-107316. The research projects funded by Natural Science Foundation of Shandong Province (ZR2017LB028), Key R&D Program of Shandong Province (2018GSF118032), and Fundamental Research Funds for the Central Universities (18CX02125A) in China are also acknowledged. M. Tinoco thanks FPU scholarship program from Ministry of Science and Innovation of Spain. H. Pan is grateful for financial support from Chinese Scholarship Council to accomplish her PhD study in University of Cadiz (Spain). TEM/STEM data were acquired at the DME-UCA node of the Unique Spanish Infrastructure for Electron Microscopy of Materials (ICTS-ELECMI).

https://dx.doi.org/10.1039/d0ta10757c.

[37] Q. Zhao, Q. Fang, H. Liu, Y. Li, H. Cui, B. Zhang, S. Tian, Halide-specific enhancement of photodegradation for sulfadiazine in estuarine waters: Roles of halogen radicals and main water constituents, Water Res. 160 (2019) 209-216. https://dx.doi.org/10.1016/j.watres.2019.05.061.

[38] N.A. Lima, G.C. Mendonça, G.T.S.T. da Silva, B. S. de Lima, E. C. Paris, M. I. B. Bernardi, Influence of the synthesis method on $CuWO_4$ nanoparticles for photocatalytic application, J. Mater. Sci. Mater. Electron. 32 (2021) 1139-1149. https://dx.doi.org/10.1007/s10854-020-04887-2.

[39] D. Wu, S. Yue, W. Wang, T. An, G. Li, H. Y. Yip, H. Zhao, P. K. Wong, Boron doped BiOBr nanosheets with enhanced photocatalytic inactivation of escherichia coli, Appl. Catal. B. 192 (2016) 35-45. https://dx.doi.org/10.1016/j.apcatb.2016.03.046.

[40] W. He, H. Jia, W.G. Wamer, Z. Zheng, P. Li, J. H. Callahan, J. Yin, Predicting and identifying reactive oxygen species and electrons for photocatalytic metal sulfide micro-nano structures, J. Catal. 320 (2014) 97-105. https://dx.doi.org/10.1016/j.jcat.2014.10.004.

[41] J.A. Herron, J. Kim, A.A. Upadhye, G. W. Huber, C. T. Maravelias, A general framework for the assessment of solar fuel technologies, Energy Environ. Sci. 8 (2015) 126-157. https://dx.doi.org/10.1039/c4ee01958j.

[42] T. Matsunaga, T. Kondo, I. Shitanda, Y. Hoshi, M. Itagaki, T. Tojo, M. Yuasa, Sensitive electrochemical detection of L-cysteine at a screen-printed diamond electrode, Carbon. 173 (2021) 395-402. https://dx.doi.org/10.1016/j.carbon.2020.10.096.

[43] S. Hashmi, M. Singh, P. Weerathunge, E. L. H. Mayes, P. D. Mariathomas, S. N. Prasad, R. Ramanathan, V. Bansal, Cobalt sulfide nanosheets as peroxidase mimics for colorimetric detection of L-cysteine, ACS Appl. Nano Mater. 4 (2021) 13352-13362. https://dx.doi.org/10.1021/acsanm.1c02851

[44] R. Prasetia, S. Fuangswasdi, F. Unob. Silver nanoparticle-supported hydroxyapatite as a material for visual detection of urinary cysteine, Anal. Methods. 11 (2019) 2888-2894. https://dx.doi.org/10.1039/c9ay00725c

[45] A. Bhatnagar, E. Kumar, M. Sillanpää. Fluoride removal from water by adsorption-a review, Chem. Eng. J. 171 (2011) 811-840. https://dx.doi.org/10.1016/j.cej.2011.05.028.

[46] O. Barbier, L. Arreola-Mendoza, L.M. Del Razo, Molecular mechanisms of fluoride toxicity, Chem-Biol. Interact. 188 (2010) 319-333. https://dx.doi.org/10.1016/j.cbi.2010.07.011.

[47] B. Liu, Z. Huang, J. Liu, Boosting the oxidase mimicking activity of nanoceria by fluoride capping: Rivaling protein enzymes and ultrasensitive F(-) detection, Nanoscale. 8 (28) (2016) 13562-7. https://dx.doi.org/10.1039/c6nr02730j.

[48] Y. Wang, T. Liu, J. Liu, Synergistically boosted degradation of organic dyes by $CeO_2$ nanoparticles with fluoride at low pH, ACS Appl. Nano Mater. 3 (2020) 842-849. https://dx.doi.org/10.1021/acsanm.9b02356.
39